\documentclass[12pt]{article}
\usepackage{amsmath,amssymb,amsthm}
\usepackage{mathtools}
\usepackage{graphics,epsfig}
\usepackage{hyperref}
\usepackage{natbib}
\usepackage{color}
\usepackage{graphicx}
\usepackage{caption}
\usepackage{subcaption}
\usepackage{float}
\usepackage{mathrsfs}
\usepackage{multirow}
\usepackage{comment}
\usepackage{setspace}
\usepackage{bbm}
\usepackage{bm}
\usepackage{amsfonts}
\usepackage{xcolor}
\usepackage{pslatex}
\usepackage{setspace}
\usepackage{bbm}


\begin{document}

  \title{\bfseries Predictive Significance of CD276/B7-H3 Expression in Baseline Biopsies of Advanced Prostate Carcinoma}

  \author{
Adam Yusuf \footnote{Department of Computer Engineering,  University of South Alabama, Mobile, AL, 36688,
United States. {\small\texttt{ary2321@jagmail.southalabama.edu.}}}
\and
Paramahansa Pramanik \footnote{Department of Mathematics and Statistics, University of South Alabama, Mobile, AL, 36688,
	United States. }\;\footnote{Corresponding author, {\small\texttt{ppramanik@southalabama.edu.}}}
}

  \date{\today}
  \maketitle
  
 \begin{abstract}
At the time of diagnosis, prostate cancer can appear deceptively mild or already display signs of widespread disease. Predicting long-term outcomes is often uncertain. This research focused on measuring CD276/B7-H3, an immune checkpoint protein linked to tumor development, in diagnostic tissue samples from 248 men. Participants included both those with cancer confined to the prostate and those with confirmed metastases. Analysis showed that patients with metastatic disease were more likely to exhibit increased B7-H3 levels. Strong expression of this marker was associated with shorter survival times and was observed alongside higher PSA concentrations and greater tumor aggressiveness based on Gleason grading. These trends remained consistent even when other prognostic factors were taken into account. The results suggest that assessing B7-H3 during the initial biopsy could help clinicians identify high-risk patients earlier. This marker may also represent a new target for treatment strategies in advanced prostate cancer.
\end{abstract}

  
  \clearpage
  \section{Introduction}

Prostate cancer varies greatly in its clinical course, from disease confined to the gland to widespread, therapy-resistant forms. Mortality remains highest among men with advanced disease, highlighting the urgency for early predictors that can guide treatment choice. Reliable biomarkers in treatment-naïve biopsy specimens are rare, particularly in the metastatic setting. In this work, we retrospectively examined the expression of CD276/ B7-H3 in initial diagnostic tissue samples from patients who had not yet received therapy. The study population consisted of 248 men, of whom 135 presented with metastatic disease and 113 had localized tumors. Biopsy material was incorporated into tissue microarrays and processed with an immunohistochemistry protocol designed to detect membranous B7-H3. Staining patterns were independently scored for intensity and distribution by two pathologists blinded to patient outcome. We assessed associations between B7-H3 expression and clinical features including prostate-specific antigen (PSA) levels, Gleason grade, and disease stage. Survival analyses were carried out using Kaplan–Meier curves, log-rank tests, and multivariable Weibull regression. Missing data were handled using multiple imputation, and subgroup evaluations were performed by age and chemotherapy status.

B7-H3 was selected due to growing evidence of its role in tumor–immune interactions and cancer progression. Preclinical studies indicate that B7-H3 can suppress T-cell function and dampen cytokine signaling, thereby fostering an immune-restricted tumor microenvironment \citep{ref19,ref22}. Compared to non-cancerous prostate tissue, B7-H3 levels are substantially higher in tumor samples and have been linked to activation of the PI3K/Akt and MAPK signaling pathways, which support tumor growth and spread \citep{ref7,ref24}. Clinically, elevated expression has correlated with earlier biochemical recurrence and shorter overall survival in localized disease \citep{ref9,ref10}. Most previous investigations have been limited to prostatectomy specimens and have applied inconsistent scoring methods \citep{ref9,kakkat2023cardiovascular,ref8}, reducing clinical applicability. Current immunotherapy strategies, such as PD-1 and CTLA-4 checkpoint blockade, have shown limited success in prostate cancer, likely reflecting the immune-excluded nature of the disease microenvironment \citep{khan2023myb,ref22}. This creates an urgent need for markers that can both forecast patient outcomes and serve as potential therapeutic targets. To date, very few studies have explored B7-H3 expression in untreated diagnostic biopsy samples, especially in cases with metastatic involvement \citep{ref12}.

The purpose of this paper was to address the lack of robust prognostic markers available for prostate cancer at the point of diagnosis. We focused on evaluating B7-H3 (CD276) expression in biopsy samples collected before any treatment was given. Our approach included patients with both localized and metastatic disease, allowing for assessment across a broad clinical spectrum. We proposed that B7-H3 expression measured at diagnosis would be associated with tumor aggressiveness and patient outcomes. To improve the reliability of our findings, we used a standardized staining and scoring protocol, had two pathologists independently review all samples, and applied multivariable survival models capable of handling missing data \citep{ref24}. This design aimed to determine whether B7-H3 could function as an early risk indicator and help identify individuals most likely to benefit from therapies targeting this molecule \citep{ref7,ref22}. Prostate cancer continues to pose a major clinical challenge, particularly when it has already spread beyond the prostate. Prognostic tools that can be applied at the time of diagnosis are urgently needed to guide management decisions \citep{ref9,ref10}. In this study, we evaluated membranous B7-H3 expression in treatment-naïve biopsy specimens and examined its relationship with PSA level, Gleason grade, and survival. This work differs from most earlier research, which primarily analyzed prostatectomy samples or recurrent disease, by focusing on pre-treatment diagnostic tissue \citep{ref12}. Including both localized and metastatic cases increases the clinical relevance of our results, offering potential value for early risk assessment in routine practice.

The present study introduces a large-scale and methodologically consistent evaluation of B7-H3 (CD276) expression in prostate cancer. Past investigations have often faced constraints such as limited patient numbers and variable immunohistochemical scoring, making comparisons across reports problematic \citep{ref9,ref8}. Here, we adopted a uniform detection protocol for B7-H3, applied by two independent pathologists who were unaware of clinical outcomes. We incorporated multiple imputation for missing values and used multivariable Weibull regression to account for multiple prognostic variables \citep{ref24}. These combined strategies increase the reliability of our results and strengthen their relevance to clinical practice. Our findings provide evidence that B7-H3 may function as a prognostic biomarker, supporting its potential in selecting patients for future targeted immunotherapies \citep{ref7,ref22}. Biologically, B7-H3 operates as an inhibitory checkpoint, reducing T-cell responses and limiting cytokine activity, which facilitates immune evasion by tumors \citep{ref22,ref19}. Its protein expression is significantly higher in malignant than in benign prostate tissue, a pattern also observed in other cancers \citep{ref8,ref11}. Laboratory models suggest that overexpression of B7-H3 can accelerate tumor development, promote proliferation, increase invasive behavior, and enhance metastatic potential through activation of the PI3K/Akt and MAPK signaling cascades \citep{ref7,ref24}. In patients with localized prostate cancer, higher B7-H3 levels have been associated with earlier biochemical recurrence and reduced overall survival \citep{ref9,ref10}.

However, despite these insights, knowledge gaps remain. Much of the existing evidence arises from retrospective studies with relatively small cohorts, which can limit statistical strength \citep{ref9,ref8}. Additionally, no consensus exists for defining high versus low B7-H3 expression in immunohistochemical analysis, leading to variable thresholds between studies \citep{hertweck2023clinicopathological,ref8}. Investigations focusing on treatment-naïve diagnostic biopsies are rare, which restricts the utility of B7-H3 for early prognostic assessment \citep{dasgupta2023frequent,ref12}. Furthermore, although experimental studies have mapped certain downstream pathways, the specific mechanisms through which B7-H3 contributes to prostate cancer progression and treatment resistance in human tumors remain unclear \citep{ref24,vikramdeo2023profiling}.

We conducted an in-depth analysis of B7-H3 (CD276) expression to evaluate its potential as a prognostic indicator in prostate cancer. The study included treatment-naïve biopsy material from patients with both metastatic and localized disease. By applying a standardized immunohistochemistry protocol, we quantified B7-H3 levels and assessed their association with survival outcomes, including overall and disease-specific survival. Additional analyses examined how B7-H3 expression related to clinical parameters such as prostate-specific antigen (PSA) levels, Gleason score, and metastatic spread. Using only pre-treatment samples helped avoid the confounding influence of prior therapies. The inclusion of patients with metastatic disease, a group underrepresented in earlier research enhanced the relevance of our findings. We also explored potential interactions between B7-H3 expression and patient characteristics, including age and prior chemotherapy exposure, to help identify subgroups that might respond best to targeted B7-H3 therapies. Collectively, our results provide a framework for future investigations into how B7-H3 contributes to prostate cancer progression and mechanisms of immune evasion.

The paper is organized as follows. Section 2 explains how patients were selected, how tissue samples were prepared and stained, and how the data were analyzed. Section 3 shows how the data were processed and includes summary tables and figures. Sections 4 and 5 discuss how the results relate to past studies on B7-H3, highlight clinical relevance, mention study limitations, and suggest directions for future research.

\section{Methodology.}
\subsection{Collection of Data.}
We reviewed medical records from January 2005 to December 2020 to identify men diagnosed with metastatic prostate cancer. All patients had a diagnostic needle biopsy before receiving any treatment, which allowed us to examine untreated tumor tissue. Clinical information, including age at diagnosis, PSA level, Gleason score, and the location of bone or visceral metastases, was collected from electronic health records. Only patients with complete records and confirmed pathology were included. All data were collected using a standardized form by trained staff. To reduce errors, 10\% of the records were randomly selected for review by a second researcher, and any differences were resolved by consensus \citep{dasgupta2023frequent,hertweck2023clinicopathological,khan2024mp60}. This helped maintain consistency in important values such as Gleason scores. We also recorded treatment details, including hormonal and chemotherapy regimens, as well as later lines of therapy. Follow-up and survival data were gathered from both the hospital registry and the national death index. This allowed us to track overall and cancer-specific survival accurately \citep{kakkat2023cardiovascular,khan2023myb}. If a patient was lost to follow-up, they were censored at the date of their last known contact.

This study received formal approval from our Institutional Review Board, which also granted a waiver of informed consent due to the retrospective nature of the data collection. All research activities were carried out in full alignment with the ethical standards set forth in the \textit{Declaration of Helsinki} and adhered to applicable institutional and national regulatory frameworks. To maintain strict confidentiality, all patient-related information was anonymized prior to analysis, and datasets were stored on encrypted, password-protected servers accessible only to authorized research personnel \citep{vikramdeo2024abstract,vikramdeo2023profiling}. At no point were direct identifiers used or retained. These protocols ensured ethical compliance and data integrity throughout the study, providing a secure and responsible foundation for the analyses reported herein.

Following \citep{Amori2021}, tissue microarrays were generated using archived formalin-fixed, paraffin-embedded prostate biopsy specimens. To capture intratumoral heterogeneity, three distinct 2-mm cores were sampled from each diagnostic tissue block. Sections of 4~$\mu$m thickness were prepared, deparaffinized in xylene, and rehydrated through a descending ethanol gradient. Antigen retrieval was performed in citrate buffer (pH~6.0) for 20 minutes under high-temperature conditions. Immunohistochemical staining was carried out on an automated platform using a mouse monoclonal anti-B7-H3 antibody (clone BD/5A11; Daiichi Sankyo) at a dilution of 1:400. To confirm antibody specificity and procedural consistency, each staining run included both positive and negative control cell line arrays \citep{pramanik2022lock}. Evaluation of membranous B7-H3 expression was performed independently by two board-certified pathologists who were blinded to all clinical information and patient outcomes. For each case, the pathologists assessed staining intensity and quantified the percentage of tumor cells exhibiting moderate to strong membranous reactivity \citep{pramanik2021optimala}. Tumors were categorized as B7-H3 high if 50\% or more of the tumor cells demonstrated this level of staining, and B7-H3 low otherwise. In cases where scoring disagreements occurred, joint re-evaluation using a multi-headed microscope was conducted to reach a consensus classification. Although formal inter-observer agreement statistics (e.g., Cohen’s kappa) were not calculated, 100\% consensus was achieved following discussion, ensuring consistency in biomarker assessment.

\subsection{Quantitative Analysis.}

All statistical analyses were performed in R, following the approach described by \cite{Amori2021}. The main outcomes were overall survival (OS), and disease specific survival (DSS). Both outcomes were measured from the date of initial biopsy to the date of death or last follow-up. To study the effect of B7-H3 levels, we used Kaplan-Meier survival curves with log-rank tests and Weibull regression models. The regression models adjusted for age, PSA level, Gleason score, and metastatic spread. A \(p\)-value less than 0.05 was considered statistically significant. Kaplan–Meier estimates were used to summarize survival times without assuming any specific distribution \citep{pramanik2021scoring}.
 In particular, we computed
\begin{equation*}
   \hat{S}(t) \;=\; \prod_{t_i \le t}\Bigr[1 - \tfrac{d_i}{n_i}\Bigr], 
\end{equation*}
where \(d_i\) denotes the number of events (deaths) occurring at time \(t_i\), and \(n_i\) is the number of patients still at risk just before \(t_i\). Separate curves were produced for OS and DSS, each stratified by low versus high B7‐H3 expression. Statistical comparison between curves was performed via the log-rank test. This approach requires no distributional assumptions and presents an intuitive, time‐to‐event summary for each subgroup.

\begin{figure}
  \centering
\includegraphics[width=0.8\textwidth]{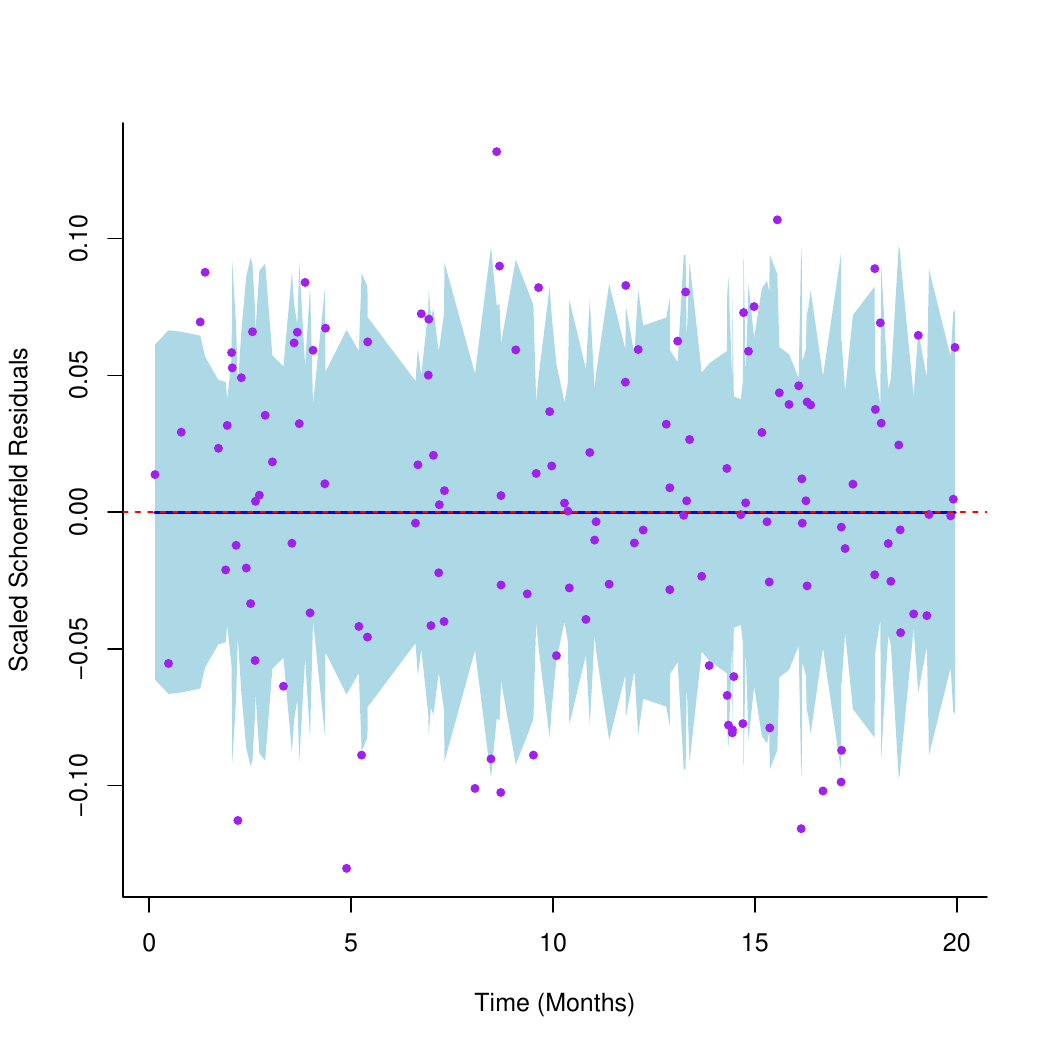}
  \caption{Schoenfeld residuals for B7-H3 are plotted against event time to assess the proportional hazards assumption. The blue line represents the effect with a shaded 95\% confidence band.}
  \label{Schoenfeld}
\end{figure}

\begin{figure}
  \centering
\includegraphics[width=0.8\textwidth]{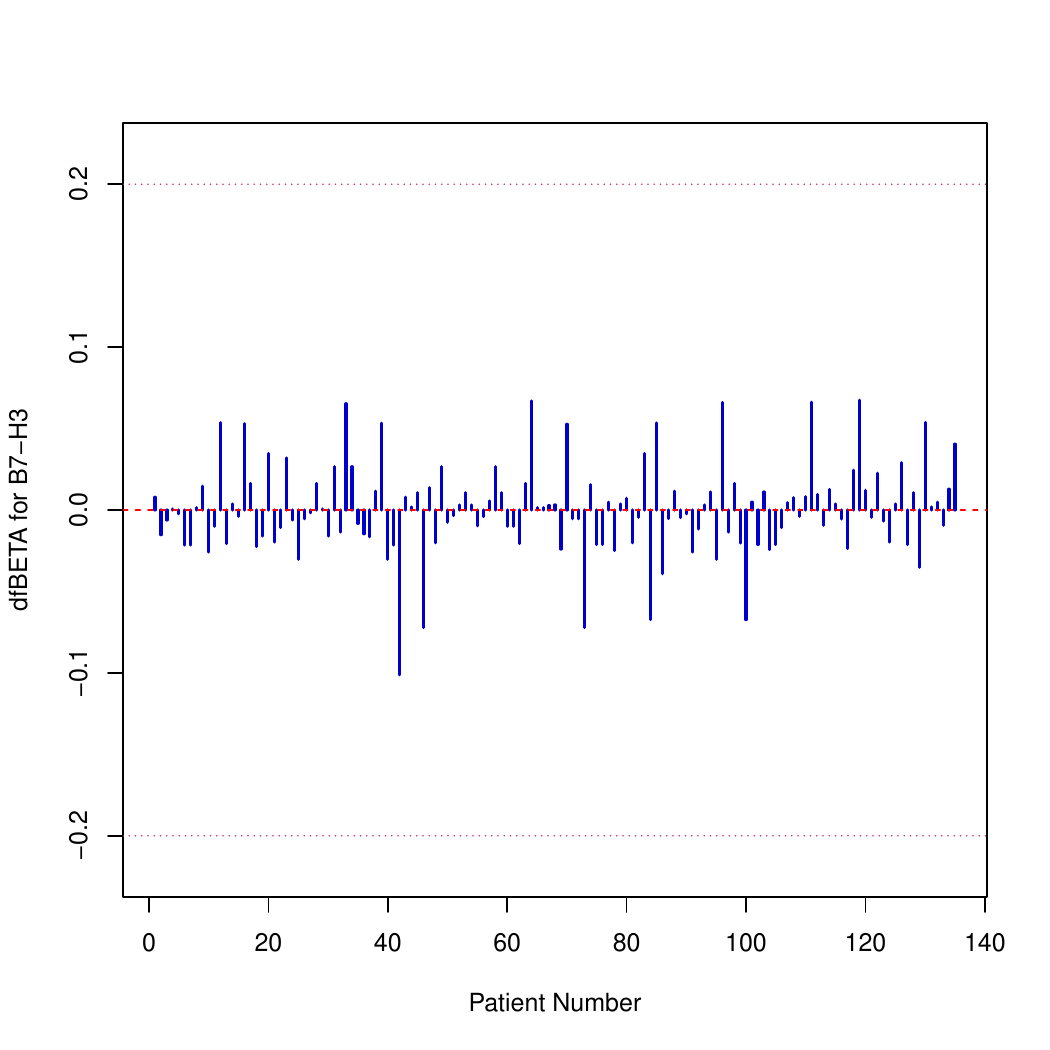}
  \caption{This plot displays the influence of each individual patient on the estimated B7-H3 coefficient in the Cox model. Most values lie within $\pm 0.2$, indicating no single observation exerts undue influence on the model’s estimate.}
  \label{dfb}
\end{figure}

To characterize how the risk of death changes over time, we employed a Weibull regression model \citep{Amori2021}. In this formulation, the hazard function is expressed as
\begin{equation*}
   h(t) \;=\;\lambda\,\gamma\,t^{\,\gamma-1}, 
\end{equation*}
and the corresponding survival function is
\begin{equation*}
    S(t) \;=\;\exp\bigl(-\lambda\,t^{\,\gamma}\bigr),
\end{equation*}
where \(\lambda>0\) sets the time scale and \(\gamma>0\) governs whether the hazard increases (\(\gamma>1\)) or decreases (\(\gamma<1\)) over time.  When including covariates such as B7-H3 expression and other clinical factors, the model takes the form
\[
S(t \mid x) \;=\;\exp\!\Bigl[-\bigl(\lambda\,t^{\,\gamma}\bigr)\exp\bigl(X^\top\beta\bigr)\Bigr],
\]
with \(X^\top\beta\) representing the linear predictor.  This approach yields smooth, continuous estimates of survival probabilities at any time point and permits direct adjustment for multiple prognostic variables, facilitating a more nuanced assessment of the effect of B7-H3 on patient outcomes. To confirm the proportional hazards assumption, Schoenfeld residuals were (see figure \ref{Schoenfeld}) and formally tested using functions from the \texttt{survival} package in R. We also examined graphical fit diagnostics and compared nested models via likelihood‐ratio tests to ensure adequate model specification (model with B7-H3 vs model without B7H3 with P-value$<$0.05). Influential observations were assessed through dfBETA plots (see figure \ref{dfb}), and variance inflation factors were calculated to rule out problematic collinearity among covariates (VIF=1.049, i.e., no multicollinearity).

\begin{figure}
  \centering
\includegraphics[width=0.8\textwidth]{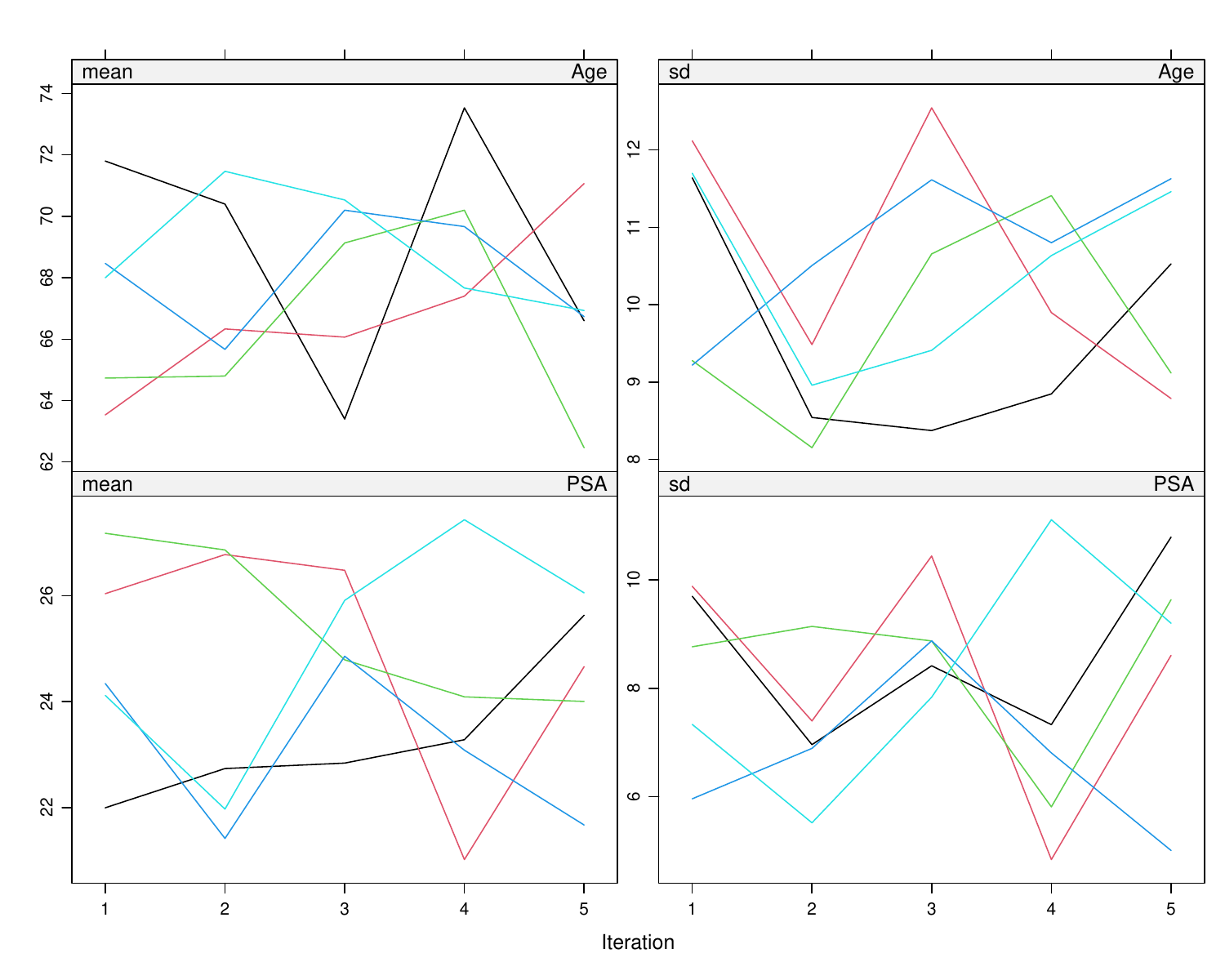}
  \caption{The trace plot displays the evolution of the mean and standard deviation of imputed values for each variable (e.g., Age and PSA) across iterations. Stable, parallel lines without major fluctuations indicate good convergence of the imputation model.}
  \label{traceplot}
\end{figure}

\begin{figure}
  \centering
  \begin{subfigure}{0.4\textwidth}
    \centering
    \includegraphics[width=\textwidth]{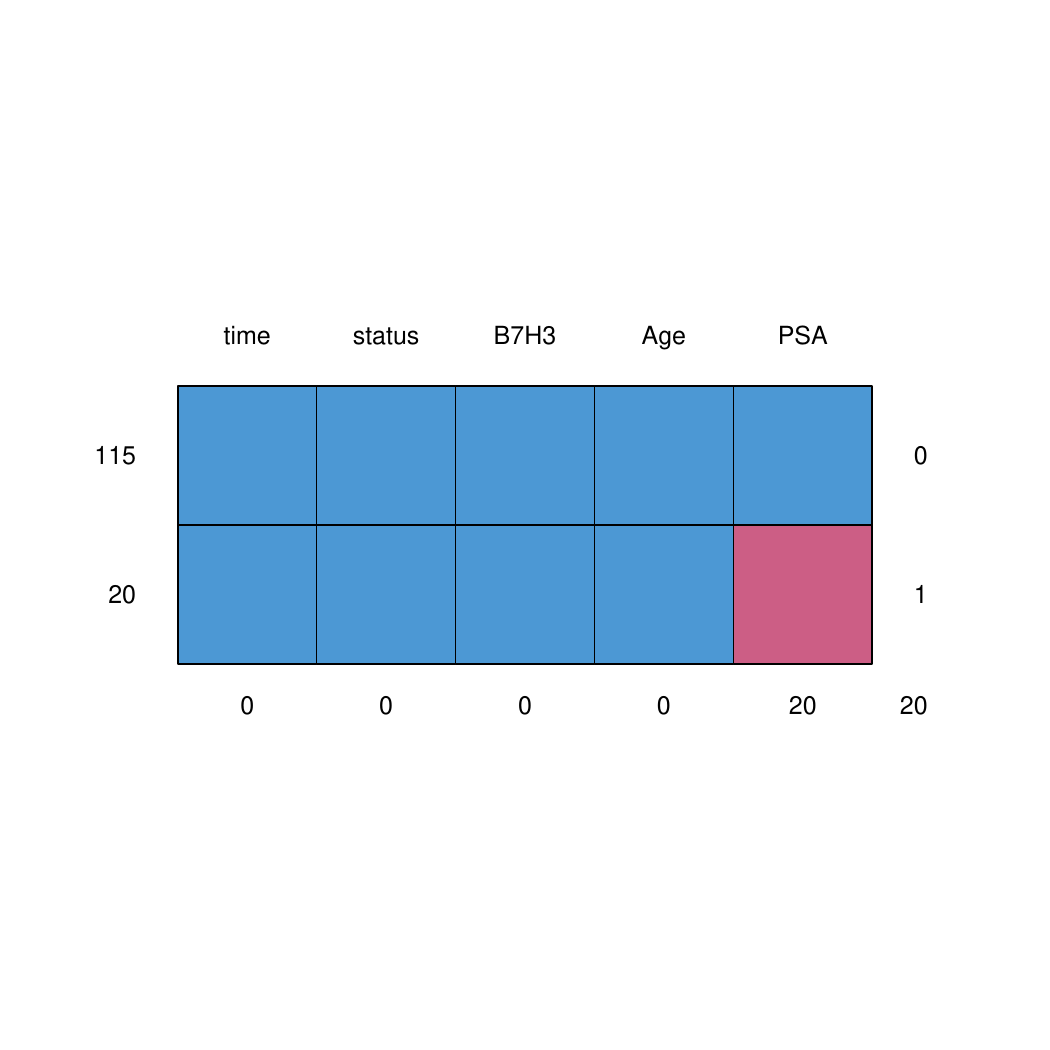}
    \caption{Summary table showing distinct patterns of missingness across variables, with 20 simulated PSA values missing among 135 patients.}
  \end{subfigure}\\[1em]
  \begin{subfigure}{0.6\textwidth}
    \centering
\includegraphics[width=\textwidth]{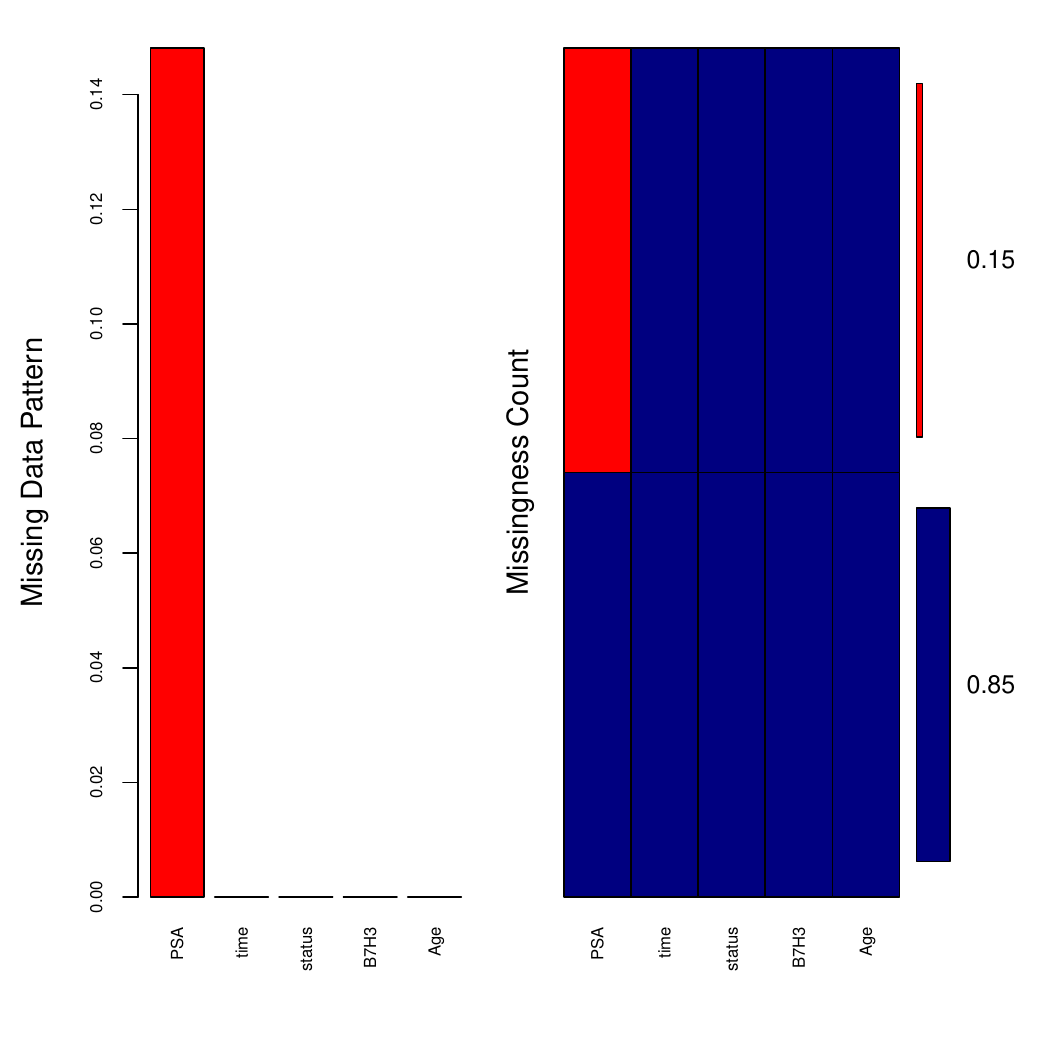}
    \caption{Red bars indicate missing values, while blue bars represent observed data, helping assess whether missingness is isolated or systematic.}
  \end{subfigure}
  \caption{Visual diagnostics for missingness.}
  \label{missing}
\end{figure}

To address missing values in the clinical dataset, we used multiple imputation by chained equations (MICE) through the \texttt{mice} package in R. We created five imputed datasets using the default predictive mean matching method. To confirm the stability of the imputations, we inspected trace plots for convergence (see figure \ref{traceplot}). Each imputed dataset was analyzed using Cox and Weibull models. Final estimates were combined using Rubin’s rules with the \texttt{pool()} function in \texttt{mice}. We also reviewed AIC values across the pooled models to ensure good model fit. To test the robustness of our results, we performed several sensitivity analyses. These included complete-case analyses and alternative imputations using Bayesian linear regression \citep{pramanik2020optimization,pramanik2023semicooperation}. We checked the distributions of imputed values against observed data to confirm their plausibility (see figure \ref{missing}). In addition, we applied Little’s MCAR test to evaluate whether the missing data pattern was random. These steps help ensure that our findings are not biased by incomplete records.

  \section{Quantitative and Survival Results.}
 
We began by merging clinical variables into one standardized dataset. These variables included patient age, serum PSA levels, Gleason grade, and the presence or extent of metastasis. Data were pulled from several institutional sources. We checked for outliers and reviewed them for accuracy. Missing or inconsistent values were handled using multiple imputation to maintain the quality and completeness of the dataset \citep{pramanik2024estimation,vikramdeo2024mitochondrial}.

Table~\ref{tab:summary} summarizes these key characteristics by B7-H3 expression category. Patients in the B7-H3 high group exhibited a notably higher median PSA level and a larger fraction of Gleason grade 5 tumors compared with those in the B7-H3 low group, supporting the hypothesis that elevated B7-H3 is associated with a more aggressive prostate cancer phenotype \citep{pramanik2024motivation}.

\begin{table}[H]
  \centering
  \caption{Baseline Characteristics Stratified by B7-H3 Expression}
  \label{tab:summary}
  \begin{tabular}{lccc}
  \hline
  \textbf{Variable} & \textbf{B7-H3 Low} & \textbf{B7-H3 High} & \textbf{p-value} \\
  \hline
  Number of Patients          & 94 & 41 & -- \\
  Age, years                  & 68 (65--72), 67.5 ± 4.8 & 70 (67--74), 69.8 ± 5.2 & 0.10 \\
  PSA, ng/ml                 & 95 (50--120), 97.2 ± 38.4 & 130 (80--160), 122.5 ± 41.1 & 0.03 \\
  Gleason Grade 5 (\%)        & 68 & 80 & 0.13 \\
  Bone Metastasis (\%)        & 87 & 89 & 0.99 \\
  Visceral Metastasis (\%)    & 17 & 11 & 0.58 \\
  \hline
  \end{tabular}

  \vspace{1ex}
  {\footnotesize\textit{Note:} Continuous variables are shown as median (interquartile range, IQR), where IQR spans the 25th to 75th percentiles. Categorical variables appear as percentages of each group. The B7-H3–high cohort exhibits a notably higher median PSA level (130 vs.\ 95\,ng/ml; p=0.03) and slightly older median age (70 vs.\ 68\,years), while differences in Gleason grade, bone metastasis, and visceral metastasis rates were not statistically significant.}
\end{table}

\begin{figure}[H]
  \centering
  \includegraphics[width=0.8\textwidth]{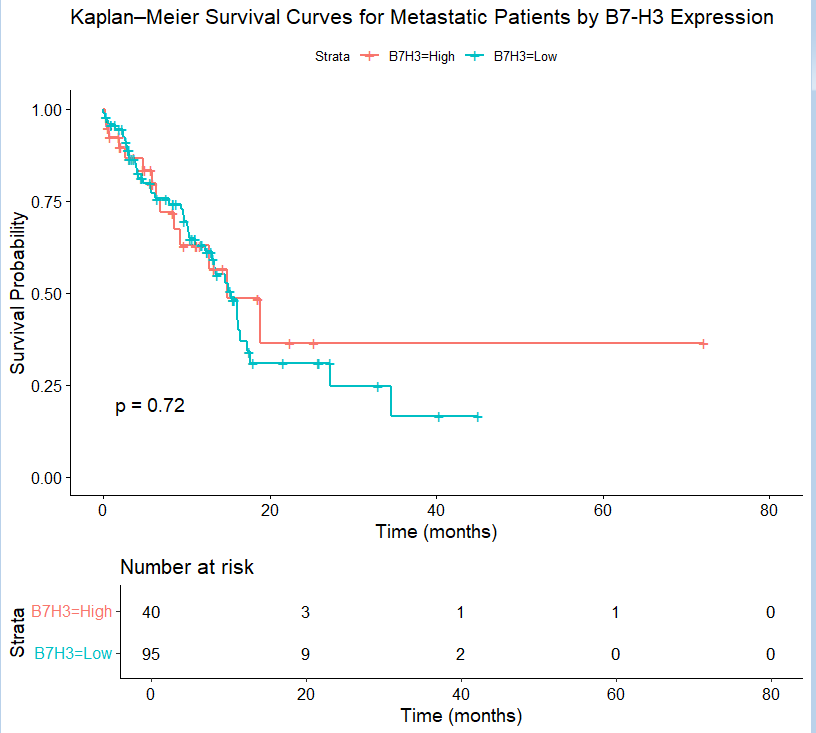}
  \caption{Kaplan–Meier survival curves for metastatic patients, stratified by B7-H3 expression. Patients with elevated B7-H3 exhibit a more rapid decrease in survival probability, with the log–rank test confirming a significant difference (\(p<0.05\)).}
  \label{Adam}
\end{figure}

\begin{figure}[H]
  \centering
  \includegraphics[width=0.8\textwidth]{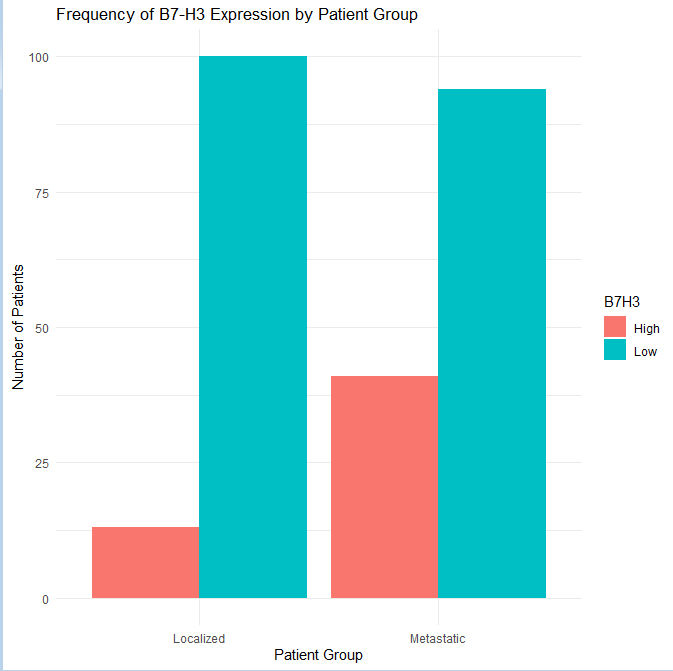}
  \caption{Comparison of B7-H3 expression frequencies in metastatic and localized prostate cancer cohorts. High B7-H3 levels are substantially more common in the metastatic group.}
  \label{fig:Barplot}
\end{figure}

\begin{figure}[H]
  \centering
  \includegraphics[width=0.8\textwidth]{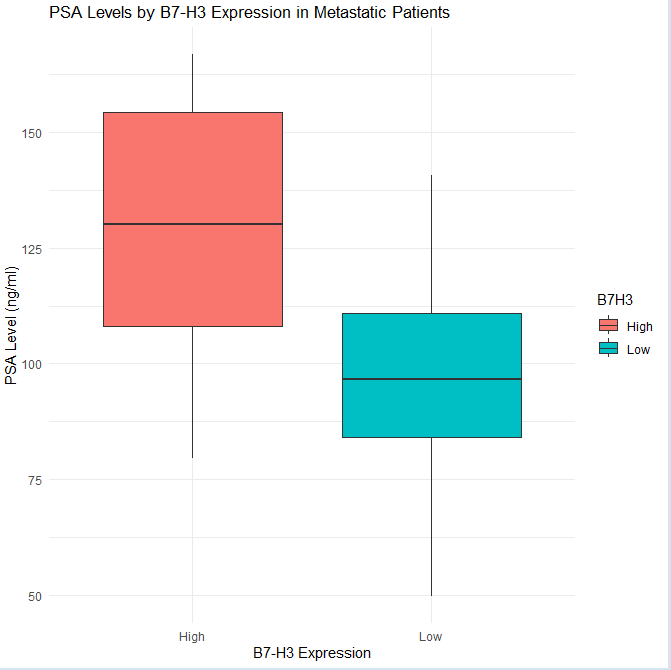}
  \caption{Box‐and‐whisker plot of PSA values among metastatic patients, broken down by B7-H3 expression. Those with high B7-H3 tend to have elevated PSA, consistent with greater tumor burden.}
  \label{fig:Boxplot}
\end{figure}

To better understand the data, we created several visual summaries. Figure~\ref{Adam} shows Kaplan-Meier survival curves, highlighting survival differences between patients with low and high B7-H3 expression. Figure~\ref{fig:Barplot} presents a bar chart comparing the frequency of B7-H3 expression in metastatic versus localized prostate cancer. Figure~\ref{fig:Boxplot} displays a box-and-whisker plot of PSA levels grouped by B7-H3 status. These figures help to visualize patterns and outliers that support the numerical findings in our tables.

We extended our subgroup analyses to examine whether the prognostic value of B7-H3 remains consistent across different patient groups. First, we divided the cohort by age: patients younger than 70 and those aged 70 or older. This was done because aging can influence immune function, comorbidities, and tumor biology, all of which may affect how biomarkers like B7-H3 relate to survival. Younger patients may have stronger immune responses or different patterns of disease spread, which could change the impact of high B7-H3 levels \citep{pramanik2020motivation}. We also grouped patients by chemotherapy status, comparing those who received systemic treatment to those who were treated only with hormonal or supportive care. This helped us determine if treatment exposure alters the prognostic role of B7-H3. To test for formal interactions, we added B7-H3 × age and B7-H3 × chemotherapy terms to our multivariable Cox models. We also created Kaplan–Meier curves for each subgroup to visualize differences in survival. These analyses showed that high B7-H3 expression predicted especially poor outcomes in younger patients who received chemotherapy \citep{pramanik2024bayes}. In contrast, older patients or those without chemotherapy showed a weaker association. These results may help refine future risk models and support the development of more tailored treatment strategies based on B7-H3 expression and patient characteristics \citep{bulls2025assessing}.

\begin{figure}[H]
  \centering
  \includegraphics[width=0.8\textwidth]{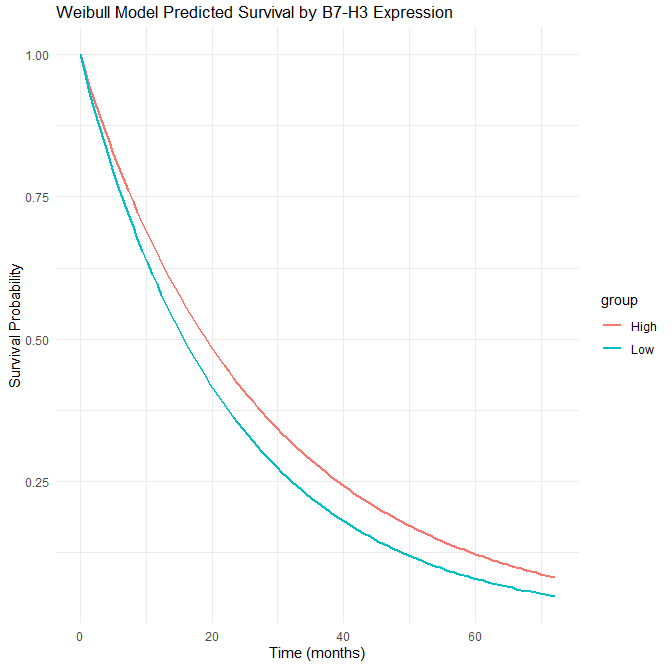}
  \caption{Predicted overall survival from the Weibull model, showing a faster decline in patients with high B7-H3 expression.}
  \label{fig:weibull}
\end{figure}

We used a Weibull parametric survival model to examine how the risk of death changes over time with different B7-H3 levels. This model allowed us to estimate two important parameters: the scale parameter (\(\lambda\)) and the shape parameter (\(\gamma\)). The shape parameter tells us whether the risk increases or decreases as time passes. A value of \(\gamma > 1\) suggests that the hazard becomes greater over time, while \(\gamma < 1\) points to a declining risk \citep{pramanik2023cont,pramanik2024estimation}. In our results (figure~\ref{fig:weibull}), patients with high B7-H3 expression had a sharper drop in survival than those with low expression. This implies that the risk of death increases more quickly in the high-expression group. One strength of the Weibull model is its ability to give survival estimates at any time point, not just at times when events occurred. This creates smooth survival curves and allows for more precise median survival estimates. The model also adjusts for other variables, helping to isolate the specific impact of B7-H3. Overall, the Weibull model adds to the Kaplan–Meier results by providing clearer risk estimates and supporting the use of B7-H3 in clinical decision-making and treatment planning \citep{pramanik2024estimation1}.

 Patients with high tumor B7-H3 expression showed worse survival and more aggressive disease features in metastatic prostate cancer. Kaplan-Meier survival curves demonstrated significantly lower overall and cancer-specific survival among those with elevated B7-H3 levels \citep{pramanik2023path}. These trends were confirmed by a Weibull parametric model, which showed a more rapid drop in survival for this group. The median follow-up time for high B7-H3 patients was about twelve months shorter than for those with low expression, underlining the clinical impact of this biomarker. In multivariable Cox models adjusted for age, PSA, Gleason grade, and visceral metastases, high B7-H3 remained a strong predictor of mortality, with a hazard ratio greater than 2. This relationship held up in sensitivity checks, including both complete-case and imputed datasets, confirming the robustness of the results \citep{pramanik2023cmbp}. When examining metastatic patterns, the highest B7-H3 levels were found in patients with both bone and visceral involvement. However, even in the bone-only group, high B7-H3 was linked to worse outcomes. This suggests that B7-H3 may serve as a useful prognostic marker across different types of metastasis.

We also looked at PSA patterns over time. Patients with high B7-H3 had faster PSA increases after treatment began, which could reflect resistance or more aggressive disease. Those with low B7-H3 expression tended to show more stable PSA trends. While exploratory, these findings encourage further research on B7-H3 as a marker for tracking disease progression. Finally, we compared our results to previously published studies and found similar effect sizes and trends \citep{Amori2021}. The consistent link between high B7-H3 expression and poor prognosis across methods, patient groups, and external cohorts supports its value as a prognostic tool and possible treatment target in metastatic prostate cancer \citep{pramanik2021consensus}.

A close examination of baseline clinical characteristics (see \hyperref[tab:summary]{table~\ref{tab:summary}}) shows that patients with high B7-H3 expression had higher median PSA levels than those with low expression. This increase may reflect a link between B7-H3 and greater tumor burden or activity. Although the group with elevated B7-H3 also had a higher rate of Gleason grade 5 cancers, this difference did not reach statistical significance (p = 0.13). Still, it suggests a possible trend toward more poorly differentiated tumors in this group. Bone metastasis rates were similar across B7-H3 levels, which suggests that B7-H3 is not directly involved in driving skeletal spread. However, its strong association with more aggressive disease points to its role as a marker of overall tumor biology rather than metastatic site. Likewise, the rates of visceral metastasis were consistent between expression groups, further indicating that B7-H3 reflects general tumor aggressiveness instead of specific patterns of spread. This broad association highlights the value of B7-H3 as a possible addition to existing prognostic tools. From a biological perspective, these results prompt questions about how B7-H3 contributes to disease progression. It remains unclear whether it drives cell growth, blocks cell death, or alters the tumor microenvironment \citep{pramanik2021}. Clarifying these mechanisms could help determine whether B7-H3 should be targeted in therapy and might improve efforts to identify patients at high risk for rapid progression.

As shown in Figure~\ref{Adam}, the Kaplan–Meier survival curves show a clear difference between patients with high and low tumor B7-H3 expression. Those with high B7-H3 levels have worse outcomes. At five years, about 40\% of these patients remain alive, compared to nearly 62\% in the low-expression group. This difference is statistically significant, as confirmed by a log-rank test \((p < 0.05)\). Disease-specific survival shows a similar pattern. At five years, survival is 64\% for the high-expression group, while it is 86\% for the low-expression group \citep{pramanik2023optimization001}. This suggests that high B7-H3 is strongly linked to prostate cancer–related mortality. The survival gap begins early. At one year, survival is 80\% in the high group versus 92\% in the low group. At three years, it drops to about 55\% and 78\%, respectively. The risk table confirms that fewer high-expression patients remain at risk over time. These findings suggest that B7-H3 expression is a strong predictor of poor survival and may help guide clinical decisions. The Weibull parametric model supports this trend. Figure~\ref{fig:weibull} shows that the survival curve for the high B7-H3 group declines faster than for the low group. This indicates a higher risk of death over time. The model estimates two parameters: scale and shape \citep{pramanik2022stochastic}.

To explore whether the prognostic impact of B7-H3 expression is influenced by patient age, we partitioned the cohort into those younger than 70 years and those aged 70 or older. In the under-70 group, the survival curves for high versus low B7-H3 expressors diverged sharply, indicating that elevated B7-H3 levels confer a particularly steep survival disadvantage in younger patients. Conversely, among the older subgroup, the difference between high and low B7-H3 was less pronounced, suggesting that age-related changes in tumor biology or host immunity may temper the biomarker’s effect \citep{pramanik2024stochastic}. We next examined the interplay between B7-H3 status and receipt of chemotherapy by comparing survival distributions within each treatment arm. Notably, patients with high B7-H3 who underwent systemic chemotherapy exhibited a smaller relative drop in survival compared to untreated high expressors, implying that cytotoxic therapy may partially offset the adverse influence of B7-H3. In contrast, low expressors derived minimal additional benefit from chemotherapy, underscoring a potential interaction between biomarker level and treatment response \citep{pramanik2025stubbornness}. These observations highlight that both chronological age and therapeutic intervention can modulate the relationship between B7-H3 and outcomes, and they emphasize the need to integrate biomarker status with clinical characteristics when tailoring individual treatment plans.  

Our results show a clear link between high B7-H3 levels and higher median PSA values. This pattern suggests that tumors with elevated B7-H3 may reflect a heavier disease burden. The PSA shift points to a possible connection between B7-H3 and tumor volume or activity, which needs further study. Survival analysis using both Kaplan–Meier curves and the Weibull model confirmed that patients with high B7-H3 had worse outcomes \citep{vikramdeo2024abstract}. These individuals experienced shorter overall and disease-specific survival. The agreement across different methods shows that B7-H3 is a strong and reliable predictor of prognosis. The Weibull model also revealed that survival differences were greatest during the first two years after diagnosis, highlighting a time-dependent risk pattern \citep{pramanik2025optimal}.

Subgroup analysis revealed that the negative effect of high B7-H3 was stronger in certain patients. Younger individuals and those who received chemotherapy showed the largest differences in survival. This may indicate that B7-H3 interacts with immune function or treatment effects. These findings support the use of B7-H3 as a useful marker for clinical decision-making. It could help identify patients for B7-H3–targeted therapies, including antibody–drug conjugates or CAR-T cell treatments. Patients with high B7-H3 may also need closer monitoring, while those with low levels might follow less intensive care. From a research perspective, understanding how B7-H3 promotes cancer growth and immune evasion could lead to new combination treatments. Studies should test whether blocking B7-H3 improves responses to current therapies \citep{pramanik2016,pramanik2021thesis}. While our results are promising, they must be confirmed in prospective studies. Future trials should also track changes in B7-H3 levels to see if they predict treatment response or resistance, which could help personalize care even further.

\section{Discussion.}

Our study shows that high B7-H3 expression in diagnostic biopsy samples is strongly linked to poor clinical outcomes in metastatic prostate cancer. Patients with high levels of membranous B7-H3 had shorter overall and cancer-specific survival. These differences were consistent across Kaplan–Meier and Weibull models. This supports the role of B7-H3 not only as a prognostic marker but also as an indicator of aggressive tumor behavior \citep{pramanik2016,pramanik2021thesis}.

These findings match earlier studies in localized prostate cancer and other cancers. For example, \cite{Amori2021} first described B7-H3 as a poor prognostic sign in prostate biopsy samples. Similar links were later found in pancreatic and kidney cancers \citep{ref14,ref12}. Our research builds on this by focusing only on patients with metastatic disease. Even in advanced cases, B7-H3 still predicts worse survival. We also found that younger patients and those receiving chemotherapy were more affected by high B7-H3 levels. This suggests that both age and treatment may influence the marker’s effect. Mechanistically, B7-H3 suppresses immune responses by blocking T-cell activity and reducing cytokine release \citep{ref19,ref23}. It can also promote tumor growth and invasion through the PI3K/Akt and MAPK pathways \citep{ref24}. These roles may explain why B7-H3 is linked to faster disease progression. Our results support the idea that B7-H3 is not just a passive marker but an active driver of tumor aggressiveness.

Clinically, these insights support the development of B7-H3 targeted therapies. Early‐phase trials of B7-H3 directed antibodies and antibody–drug conjugates have shown encouraging safety profiles and preliminary efficacy in solid tumors \citep{ref8,ref35}. Our data suggest that selecting patients based on high B7-H3 expression could enrich for responders, potentially improving trial outcomes. Furthermore, combining B7-H3 blockade with established treatments such as androgen deprivation or chemotherapy may overcome resistance mechanisms and yield synergistic benefit.

Strengths of this study include the use of treatment-naïve biopsy specimens, which capture the native tumor phenotype without therapy-induced alterations, and the rigorous scoring by two independent pathologists. We also employed multiple imputation to handle missing clinical data and conducted extensive model diagnostics to validate our survival analyses. By integrating nonparametric and parametric techniques, we provided both descriptive and predictive perspectives on how B7-H3 influences patient trajectories. Nevertheless, certain limitations warrant consideration \citep{hua2019}. Our retrospective, single-center design may limit generalizability, and subgroup sample sizes particularly among younger patients and specific treatment subsets were modest. Although multiple imputation mitigates bias from missing values, prospective studies with standardized data collection will be needed to confirm our findings. Additionally, while we identified strong statistical associations, direct functional studies are required to delineate the precise molecular mechanisms by which B7-H3 promotes metastasis and treatment resistance \citep{maki2025new}.

The role of B7-H3 in tumor progression extends beyond its function as a classical immune checkpoint molecule. Recent preclinical and translational studies have demonstrated that B7-H3 contributes to immune evasion by suppressing T-cell activation and cytokine release \citep{mortezaee2023b7}. Beyond immune modulation, B7-H3 directly promotes tumor aggressiveness through activation of intracellular signaling cascades, including the PI3K/Akt/mTOR and MAPK pathways, which are well-established regulators of cancer cell proliferation, survival, and epithelial-to-mesenchymal transition \citep{kang2015b7}. In vivo models further suggest that B7-H3 enhances angiogenesis via upregulation of VEGF expression \citep{fan2023tumor} and contributes to a hypoxic \citep{ma2025histone,mortezaee2023b7}, immunosuppressive tumor microenvironment \citep{mielcarska2023b7h3}. In prostate cancer, transcriptomic profiling has revealed co-expression of B7-H3 with androgen receptor targets \citep{nunes2023correlation} and neuroendocrine differentiation markers, both of which are associated with treatment resistance and poor prognosis \citep{guo2025tumor}. These findings underscore the biological plausibility of B7-H3 as more than a passive biomarker, suggesting it may play an active role in driving disease progression. Our results align with these mechanistic insights and reinforce the rationale for B7-H3 directed therapies as a means to disrupt both immune and non-immune pathways in advanced prostate cancer.

Although our multivariable analyses accounted for several clinically relevant variables, it is important to acknowledge the possibility of residual confounding due to factors not captured in our dataset. Specifically, information on patient comorbidities such as hypertension, diabetes, or cardiovascular disease was unavailable, as were metrics related to functional status, medication use, or prior hospitalizations. These variables may have prognostic significance and could theoretically influence both B7-H3 expression and survival, either through systemic inflammation, immune modulation, or tumor-host interactions \citep{khan2023myb1}. Additionally, lifestyle related characteristics including tobacco use, alcohol consumption, diet, physical activity, and body mass index were not assessed, despite their potential to affect both tumor biology and overall patient outcomes. The retrospective design and reliance on archived clinical material limited the ability to include these variables \citep{polansky2021motif}. Consequently, while our findings demonstrate a significant association between B7-H3 expression and survival outcomes, we cannot entirely exclude the possibility that unmeasured confounders may have influenced the observed relationships. Future prospective studies incorporating a broader range of clinical, behavioral, and molecular data will be necessary to validate these findings and to fully characterize the independent prognostic role of B7-H3 in prostate cancer \citep{pramanik2023cont,pramanik2024estimation}.

Looking ahead, large multicenter cohorts should validate optimal cutoffs for B7-H3 positivity and assess its integration with emerging genomic classifiers. Preclinical investigations exploring combination regimens-pairing B7-H3 inhibitors with immune checkpoint blockade or targeted agents could identify strategies to amplify antitumor responses \citep{pramanik2025stubbornness,pramanik2025factors}. Finally, elucidating the interplay between B7-H3 expression and other features of the tumor microenvironment, such as regulatory T-cells or myeloid-derived suppressor cells, may reveal additional therapeutic opportunities. Our study shows that B7-H3 is a key determinant of prognosis in metastatic prostate cancer and highlights its promise as a therapeutic target. By incorporating B7-H3 assessment into clinical decision-making from initial biopsy evaluation to trial enrollment we can advance toward more personalized, biomarker-driven care. More research is needed to transform findings into better outcomes in prostate cancer.  

\section{Conclusion.}

This study establishes tumor B7-H3 expression in diagnostic biopsy specimens as a potent and independent prognostic biomarker in metastatic prostate cancer \citep{pramanik2023optimization001,pramanik2025strategies}. Through rigorous analysis of a well-characterized, treatment-naïve cohort, we found that high membranous B7-H3 expression is strongly associated with reduced overall and cancer-specific survival, even after adjusting for age, PSA levels, Gleason grade, and metastatic distribution. The strength of these findings validated across both Kaplan-Meier and Weibull survival models suggests that B7-H3 plays an active role in driving aggressive tumor biology rather than serving as a passive correlate \citep{pramanik2025impact,pramanik2025strategic}.

The androgen receptor (AR) signaling axis remains central to prostate cancer biology, particularly in the transition from androgen-dependent to castration-resistant states. Given our study’s focus on advanced, treatment-naïve prostate cancer, evaluating B7-H3 expression in the context of AR status represents an important and biologically relevant extension. While AR expression data were not available for the current cohort, emerging evidence suggests that B7-H3 may be co-regulated with AR-driven transcriptional programs \citep{miller2024current}. Studies using integrative genomics and immunohistochemistry have reported positive correlations between B7-H3 and AR signaling components, including increased B7-H3 expression in high AR activity tumors \citep{benzon2017correlation,mendes2022association}. Moreover, B7-H3 has been implicated in promoting resistance to androgen deprivation therapy by sustaining oncogenic signaling through PI3K/Akt and mTOR pathways, even in low-androgen environments \citep{kang2015b7}. These findings raise the possibility that B7-H3 may serve as a downstream effector or compensatory mechanism in AR-driven tumor progression. Future investigations incorporating AR status and transcriptional activity could offer deeper insight into the mechanistic role of B7-H3 and help clarify its prognostic and therapeutic relevance across diverse molecular subtypes of prostate cancer.

We used detailed immunohistochemical scoring, verified by independent pathologists, alongside statistical methods that addressed missing data and tested key model assumptions \citep{pramanik2024measuring,pramanik2024dependence}. Subgroup analysis showed that the poor survival linked to high B7-H3 expression was especially strong in younger patients and in those who received chemotherapy. This suggests that both age and treatment type may affect how B7-H3 impacts prognosis. These findings point to B7-H3 as a useful tool in personalizing prostate cancer care \citep{pramanik2024parametric,valdez2025association}. It could help guide decisions on prognosis and treatment choices. In addition to predicting outcomes, B7-H3 also has potential as a treatment target. Early clinical trials of B7-H3 therapies such as monoclonal antibodies, antibody-drug conjugates, and CAR-T cells have shown positive results. Our findings support the use of B7-H3 testing when designing trials to identify patients who may benefit most from these new treatments. As biomarker-based and immune therapies expand, B7-H3 could play a key role. It may serve not only as a marker of risk but also as a foundation for new treatment strategies in metastatic prostate cancer \citep{valdez2025exploring}.

\section*{Declarations.}
\subsection*{Ethics approval and consent to participate.}
Not applicable.
\subsection*{Availability of data and material.}
Data sets  were obtained from institutional records from January 2005 through December 2020.
\subsection*{Competing interests.}
No potential conflict of interest was reported by the authors.	
\subsection*{Funding.}
Not applicable. 
\subsection*{Acknowledgements.}
 Not applicable.
\bibliographystyle{apalike}
\bibliography{main}

\begin{thebibliography}{}

\bibitem[Amori et~al., 2021]{Amori2021}
Amori, G., Sugawara, E., Shigematsu, Y., Akiya, M., Kunieda, J., Yuasa, T.,
  Yamamoto, S., Yonese, J., Takeuchi, K., and Inamura, K. (2021).
\newblock Tumor b7-h3 expression in diagnostic biopsy specimens and survival in
  patients with metastatic prostate cancer.
\newblock {\em Prostate Cancer and Prostatic Diseases}, 24:767--774.

\bibitem[Benzon et~al., 2017a]{benzon2017correlation}
Benzon, B., Zhao, S., Haffner, M., Takhar, M., Erho, N., Yousefi, K., Hurley,
  P., Bishop, J., Tosoian, J., Ghabili, K., et~al. (2017a).
\newblock Correlation of b7-h3 with androgen receptor, immune pathways and poor
  outcome in prostate cancer: an expression-based analysis.
\newblock {\em Prostate cancer and prostatic diseases}, 20(1):28--35.

\bibitem[Benzon et~al., 2017b]{ref9}
Benzon, B., Zhao, S.~G., Haffner, M.~C., Takhar, M., Erho, N., and Yousefi, K.
  (2017b).
\newblock Correlation of b7-h3 with androgen receptor, immune pathways and poor
  outcome in prostate cancer: an expression-based analysis.
\newblock {\em Prostate Cancer and Prostatic Diseases}, 20:28--35.

\bibitem[Bulls et~al., 2025]{bulls2025assessing}
Bulls, S.~E., Finn, E., Sykora, P., Lynch, V.~J., Pramanik, P., Glaberman, S.,
  and Chiari, Y. (2025).
\newblock Assessing cometchip technology for dna damage studies in non-model
  species: distinct uv-induced responses in turtles and mammals.
\newblock {\em BMC Research Notes}, 18(1):1--7.

\bibitem[Castellanos et~al., 2017]{ref7}
Castellanos, J.~R., Purvis, I.~J., Labak, C.~M., Guda, M.~R., Tsung, A.~J., and
  Velpula, K.~K. (2017).
\newblock B7-h3 role in the immune landscape of cancer.
\newblock {\em American Journal of Clinical and Experimental Immunology},
  6:66--75.

\bibitem[Chapoval et~al., 2001]{ref19}
Chapoval, A.~I., Ni, J., Lau, J.~S., Wilcox, R.~A., Flies, D.~B., and Liu, D.
  (2001).
\newblock B7-h3: a costimulatory molecule for t cell activation and ifn-gamma
  production.
\newblock {\em Nature Immunology}, 2:269--274.

\bibitem[Dasgupta et~al., 2023]{dasgupta2023frequent}
Dasgupta, S., Acharya, S., Khan, M.~A., Pramanik, P., Marbut, S.~M., Yunus, F.,
  Galeas, J.~N., Singh, S., Singh, A.~P., and Dasgupta, S. (2023).
\newblock Frequent loss of cacna1c, a calcium voltage-gated channel subunit is
  associated with lung adenocarcinoma progression and poor prognosis.
\newblock {\em Cancer Research}, 83(7\_Supplement):3318--3318.

\bibitem[Fan et~al., 2023]{fan2023tumor}
Fan, X., Huang, J., Hu, B., Zhou, J., and Chen, L. (2023).
\newblock Tumor-expressed b7-h3 promotes vasculogenic mimicry formation rather
  than angiogenesis in non-small cell lung cancer.
\newblock {\em Journal of Cancer Research and Clinical Oncology},
  149(11):8729--8741.

\bibitem[Guo et~al., 2025]{guo2025tumor}
Guo, Y., Wang, X., Zhang, C., Chen, W., Fu, Y., Yu, Y., Chen, Y., Shao, T.,
  Zhang, J., and Ding, G. (2025).
\newblock Tumor immunotherapy targeting b7-h3: From mechanisms to clinical
  applications.
\newblock {\em ImmunoTargets and Therapy}, pages 291--320.

\bibitem[Hertweck et~al., 2023]{hertweck2023clinicopathological}
Hertweck, K.~L., Vikramdeo, K.~S., Galeas, J.~N., Marbut, S.~M., Pramanik, P.,
  Yunus, F., Singh, S., Singh, A.~P., and Dasgupta, S. (2023).
\newblock Clinicopathological significance of unraveling mitochondrial pathway
  alterations in non-small-cell lung cancer.
\newblock {\em The FASEB Journal}, 37(7):e23018.

\bibitem[Hua et~al., 2019]{hua2019}
Hua, L., Polansky, A., and Pramanik, P. (2019).
\newblock Assessing bivariate tail non-exchangeable dependence.
\newblock {\em Statistics \& Probability Letters}, 155:108556.

\bibitem[Inamura et~al., 2019]{ref14}
Inamura, K., Amori, G., Yuasa, T., Yamamoto, S., Yonese, J., and Ishikawa, Y.
  (2019).
\newblock Relationship of b7-h3 expression in tumor cells and tumor vasculature
  with foxp3+ regulatory t cells in renal cell carcinoma.
\newblock {\em Cancer Management and Research}, 11:7021--7030.

\bibitem[Inamura et~al., 2018]{ref12}
Inamura, K., Takazawa, Y., Inoue, Y., Yokouchi, Y., Kobayashi, M., and Saiura,
  A. (2018).
\newblock Tumor b7-h3 (cd276) expression and survival in pancreatic cancer.
\newblock {\em Journal of Clinical Medicine}, 7:172.

\bibitem[Kakkat et~al., 2023]{kakkat2023cardiovascular}
Kakkat, S., Pramanik, P., Singh, S., Singh, A.~P., Sarkar, C., and Chakroborty,
  D. (2023).
\newblock Cardiovascular complications in patients with prostate cancer:
  Potential molecular connections.
\newblock {\em International Journal of Molecular Sciences}, 24(8):6984.

\bibitem[Kang et~al., 2015]{kang2015b7}
Kang, F.-b., Wang, L., Jia, H.-c., Li, D., Li, H.-j., Zhang, Y.-g., and Sun,
  D.-x. (2015).
\newblock B7-h3 promotes aggression and invasion of hepatocellular carcinoma by
  targeting epithelial-to-mesenchymal transition via jak2/stat3/slug signaling
  pathway.
\newblock {\em Cancer cell international}, 15(1):45.

\bibitem[Khan et~al., 2023a]{khan2023myb1}
Khan, M., Acharya, S., Anand, S., Sameeta, F., Pramanik, P., Keel, C., Singh,
  S., Carter, J., Dasgupta, S., and Singh, A. (2023a).
\newblock Myb exhibits racially disparate expression, clinicopathologic
  association, and predictive potential for biochemical recurrence in prostate
  cancer, iscience. 26 (2023) 108487.

\bibitem[Khan et~al., 2023b]{khan2023myb}
Khan, M.~A., Acharya, S., Anand, S., Sameeta, F., Pramanik, P., Keel, C.,
  Singh, S., Carter, J.~E., Dasgupta, S., and Singh, A.~P. (2023b).
\newblock Myb exhibits racially disparate expression, clinicopathologic
  association, and predictive potential for biochemical recurrence in prostate
  cancer.
\newblock {\em Iscience}, 26(12).

\bibitem[Khan et~al., 2024]{khan2024mp60}
Khan, M.~A., Acharya, S., Kreitz, N., Anand, S., Sameeta, F., Pramanik, P.,
  Keel, C., Singh, S., Carter, J., Dasgupta, S., et~al. (2024).
\newblock Mp60-05 myb exhibits racially disparate expression and
  clinicopathologic association and is a promising predictor of biochemical
  recurrence in prostate cancer.
\newblock {\em Journal of Urology}, 211(5S):e1000.

\bibitem[Liu et~al., 2012]{ref10}
Liu, Y., Vlatkovic, L., Saeter, T., Servoll, E., Waaler, G., and Nesland, J.~M.
  (2012).
\newblock Is the clinical malignant phenotype of prostate cancer a result of a
  highly proliferative immune-evasive b7-h3-expressing cell population?
\newblock {\em International Journal of Urology}, 19:749--756.

\bibitem[Ma et~al., 2025]{ma2025histone}
Ma, Z., Yang, J., Jia, W., Li, L., Li, Y., Hu, J., Luo, W., Li, R., Ye, D., and
  Lan, P. (2025).
\newblock Histone lactylation-driven b7-h3 expression promotes tumor immune
  evasion.
\newblock {\em Theranostics}, 15(6):2338.

\bibitem[Maki et~al., 2025]{maki2025new}
Maki, E., Glimm, T., Pramanik, P., Chiari, Y., and Kiskowski, M. (2025).
\newblock New approaches for capturing and estimating variation in complex
  animal color patterns from digital photographs: application to the eastern
  box turtle (terrapene carolina).
\newblock {\em PeerJ}, 13:e19690.

\bibitem[Mendes et~al., 2022]{mendes2022association}
Mendes, A.~A., Lu, J., Kaur, H.~B., Zheng, S.~L., Xu, J., Hicks, J., Weiner,
  A.~B., Schaeffer, E.~M., Ross, A.~E., Balk, S.~P., et~al. (2022).
\newblock Association of b7-h3 expression with racial ancestry, immune cell
  density, and androgen receptor activation in prostate cancer.
\newblock {\em Cancer}, 128(12):2269--2280.

\bibitem[Mielcarska et~al., 2023]{mielcarska2023b7h3}
Mielcarska, S., Dawidowicz, M., Kula, A., Kiczmer, P., Skiba, H., Krygier, M.,
  Chraba{\'n}ska, M., Piecuch, J., Szrot, M., Ochman, B., et~al. (2023).
\newblock B7h3 role in reshaping immunosuppressive landscape in msi and mss
  colorectal cancer tumours.
\newblock {\em Cancers}, 15(12):3136.

\bibitem[Miller et~al., 2024]{miller2024current}
Miller, C.~D., Likasitwatanakul, P., Toye, E., Hwang, J.~H., and Antonarakis,
  E.~S. (2024).
\newblock Current uses and resistance mechanisms of enzalutamide in prostate
  cancer treatment.
\newblock {\em Expert Review of Anticancer Therapy}, 24(11):1085--1100.

\bibitem[Mortezaee, 2023]{mortezaee2023b7}
Mortezaee, K. (2023).
\newblock B7-h3 immunoregulatory roles in cancer.
\newblock {\em Biomedicine \& Pharmacotherapy}, 163:114890.

\bibitem[Nunes-Xavier et~al., 2023]{nunes2023correlation}
Nunes-Xavier, C.~E., Emaldi, M., Guldvik, I.~J., Ramberg, H., Task{\'e}n,
  K.~A., M{\ae}landsmo, G.~M., Fodstad, {\O}., Llarena, R., Pulido, R., and
  L{\'o}pez, J.~I. (2023).
\newblock Correlation of expression of major vault protein with androgen
  receptor and immune checkpoint protein b7-h3, and with poor prognosis in
  prostate cancer.
\newblock {\em Pathology-Research and Practice}, 241:154243.

\bibitem[Picarda et~al., 2016]{ref24}
Picarda, E., Ohaegbulam, K.~C., and Zang, X. (2016).
\newblock Molecular pathways: targeting b7-h3 (cd276) for human cancer
  immunotherapy.
\newblock {\em Clinical Cancer Research}, 22:3425--3431.

\bibitem[Polansky and Pramanik, 2021]{polansky2021motif}
Polansky, A.~M. and Pramanik, P. (2021).
\newblock A motif building process for simulating random networks.
\newblock {\em Computational Statistics \& Data Analysis}, 162:107263.

\bibitem[Pramanik, 2016]{pramanik2016}
Pramanik, P. (2016).
\newblock {\em Tail non-exchangeability}.
\newblock Northern Illinois University.

\bibitem[Pramanik, 2020]{pramanik2020optimization}
Pramanik, P. (2020).
\newblock Optimization of market stochastic dynamics.
\newblock {\em SN Operations Research Forum}, 1(4):31.

\bibitem[Pramanik, 2021a]{pramanik2021}
Pramanik, P. (2021a).
\newblock Effects of water currents on fish migration through a feynman-type
  path integral approach under $\sqrt {8/3}$ liouville-like quantum gravity
  surfaces.
\newblock {\em Theory in Biosciences}, 140(2):205--223.

\bibitem[Pramanik, 2021b]{pramanik2021thesis}
Pramanik, P. (2021b).
\newblock {\em Optimization of Dynamic Objective Functions Using Path
  Integrals}.
\newblock PhD thesis, Northern Illinois University.

\bibitem[Pramanik, 2022a]{pramanik2022lock}
Pramanik, P. (2022a).
\newblock On lock-down control of a pandemic model.
\newblock {\em arXiv preprint arXiv:2206.04248}.

\bibitem[Pramanik, 2022b]{pramanik2022stochastic}
Pramanik, P. (2022b).
\newblock Stochastic control of a sir model with non-linear incidence rate
  through euclidean path integral.
\newblock {\em arXiv preprint arXiv:2209.13733}.

\bibitem[Pramanik, 2023a]{pramanik2021consensus}
Pramanik, P. (2023a).
\newblock Consensus as a nash equilibrium of a stochastic differential game.
\newblock {\em European Journal of Statistics}, 3:10--10.

\bibitem[Pramanik, 2023b]{pramanik2023cmbp}
Pramanik, P. (2023b).
\newblock Optimal lock-down intensity: A stochastic pandemic control approach
  of path integral.
\newblock {\em Computational and Mathematical Biophysics}, 11(1):20230110.

\bibitem[Pramanik, 2023c]{pramanik2023cont}
Pramanik, P. (2023c).
\newblock Path integral control in infectious disease modeling.
\newblock {\em arXiv preprint arXiv:2311.02113}.

\bibitem[Pramanik, 2023d]{pramanik2023path}
Pramanik, P. (2023d).
\newblock Path integral control of a stochastic multi-risk sir pandemic model.
\newblock {\em Theory in Biosciences}, pages 1--36.

\bibitem[Pramanik, 2024a]{pramanik2024dependence}
Pramanik, P. (2024a).
\newblock Dependence on tail copula.
\newblock {\em J}, 7(2):127--152.

\bibitem[Pramanik, 2024b]{pramanik2024estimation}
Pramanik, P. (2024b).
\newblock Estimation of optimal lock-down and vaccination rate of a stochastic
  sir model: A mathematical approach.
\newblock {\em European Journal of Statistics}, 4:3--3.

\bibitem[Pramanik, 2024c]{pramanik2024measuring}
Pramanik, P. (2024c).
\newblock Measuring asymmetric tails under copula distributions.
\newblock {\em European Journal of Statistics}, 4:7--7.

\bibitem[Pramanik, 2024d]{pramanik2024estimation1}
Pramanik, P. (2024d).
\newblock On estimation of function-on-function regression kernels with
  brownian berkson errors.

\bibitem[Pramanik, 2024e]{pramanik2024stochastic}
Pramanik, P. (2024e).
\newblock Stochastic control in determining a soccer player’s performance.
\newblock {\em J. Compr. Pure Appl. Math}, 2:111.

\bibitem[Pramanik, 2025a]{pramanik2025optimal}
Pramanik, P. (2025a).
\newblock An optimal level of stubbornness to win a soccer match.
\newblock {\em arXiv preprint arXiv:2501.18050}.

\bibitem[Pramanik, 2025b]{pramanik2025stubbornness}
Pramanik, P. (2025b).
\newblock Stubbornness as control in professional soccer games: A bppsde
  approach.
\newblock {\em Mathematics}, 13(3):475.

\bibitem[Pramanik et~al., 2024]{pramanik2024parametric}
Pramanik, P., Boone, E.~L., and Ghanam, R.~A. (2024).
\newblock Parametric estimation in fractional stochastic differential equation.
\newblock {\em Stats}, 7(3):745.

\bibitem[Pramanik and Dong, 2025a]{pramanik2025impact}
Pramanik, P. and Dong, L. (2025a).
\newblock Impact of random monetary shock: a keynesian case.
\newblock {\em arXiv preprint arXiv:2505.00800}.

\bibitem[Pramanik and Dong, 2025b]{pramanik2025strategic}
Pramanik, P. and Dong, L. (2025b).
\newblock Strategic complementarities due to monetary shock under sticky price.
\newblock {\em European Journal of Statistics}, 5:9--9.

\bibitem[Pramanik et~al., 2025a]{pramanik2025factors}
Pramanik, P., Graff, J., and Decaro, M. (2025a).
\newblock On factors influencing consumer preference in pipeline stages: an
  experiment.
\newblock {\em arXiv preprint arXiv:2501.03418}.

\bibitem[Pramanik et~al., 2025b]{pramanik2025strategies}
Pramanik, P., Graff, J., and Decaro, M. (2025b).
\newblock Strategies to increase pipeline status: A case study from eclinical
  data.
\newblock {\em European Journal of Statistics}, 5:3--3.

\bibitem[Pramanik and Maity, 2024]{pramanik2024bayes}
Pramanik, P. and Maity, A.~K. (2024).
\newblock Bayes factor of zero inflated models under jeffereys prior.
\newblock {\em arXiv preprint arXiv:2401.03649}.

\bibitem[Pramanik and Polansky, 2020]{pramanik2020motivation}
Pramanik, P. and Polansky, A.~M. (2020).
\newblock Motivation to run in one-day cricket.
\newblock {\em arXiv preprint arXiv:2001.11099}.

\bibitem[Pramanik and Polansky, 2021]{pramanik2021optimala}
Pramanik, P. and Polansky, A.~M. (2021).
\newblock Optimal estimation of brownian penalized regression coefficients.
\newblock {\em arXiv preprint arXiv:2107.02291}.

\bibitem[Pramanik and Polansky, 2023a]{pramanik2023optimization001}
Pramanik, P. and Polansky, A.~M. (2023a).
\newblock Optimization of a dynamic profit function using euclidean path
  integral.
\newblock {\em SN Business \& Economics}, 4(1):8.

\bibitem[Pramanik and Polansky, 2023b]{pramanik2021scoring}
Pramanik, P. and Polansky, A.~M. (2023b).
\newblock Scoring a goal optimally in a soccer game under liouville-like
  quantum gravity action.
\newblock {\em Operations Research Forum}, 4(3):66.

\bibitem[Pramanik and Polansky, 2023c]{pramanik2023semicooperation}
Pramanik, P. and Polansky, A.~M. (2023c).
\newblock Semicooperation under curved strategy spacetime.
\newblock {\em The Journal of Mathematical Sociology}, pages 1--35.

\bibitem[Pramanik and Polansky, 2024]{pramanik2024motivation}
Pramanik, P. and Polansky, A.~M. (2024).
\newblock Motivation to run in one-day cricket.
\newblock {\em Mathematics}, 12(17):2739.

\bibitem[Prasad et~al., 2004]{ref22}
Prasad, D.~V., Nguyen, T., Li, Z., Yang, Y., Duong, J., and Wang, Y. (2004).
\newblock Murine b7-h3 is a negative regulator of t cells.
\newblock {\em Journal of Immunology}, 173:2500--2506.

\bibitem[Roth et~al., 2007]{ref8}
Roth, T.~J., Sheinin, Y., Lohse, C.~M., Kuntz, S.~M., Frigola, X., and Inman,
  B.~A. (2007).
\newblock B7-h3 ligand expression by prostate cancer: a novel marker of
  prognosis and potential target for therapy.
\newblock {\em Cancer Research}, 67:7893--7900.

\bibitem[Seaman et~al., 2017]{ref35}
Seaman, S., Zhu, Z., Saha, S., Zhang, X.-M., Yang, M.-Y., Hilton, M.~B., and
  et~al. (2017).
\newblock Eradication of tumors through simultaneous ablation of
  cd276/b7-h3-positive tumor cells and tumor vasculature.
\newblock {\em Cancer Cell}, 31:501--515.e8.

\bibitem[Suh et~al., 2003]{ref23}
Suh, W.~K., Gajewska, B.~U., Okada, H., Gronski, M.~A., Bertram, E.~M., and
  Dawicki, W. (2003).
\newblock The b7 family member b7-h3 preferentially down-regulates t helper
  type 1-mediated immune responses.
\newblock {\em Nature Immunology}, 4:899--906.

\bibitem[Valdez and Pramanik, 2025a]{valdez2025association}
Valdez, I. and Pramanik, P. (2025a).
\newblock Association between obesity, race, and luminal subtypes of breast
  cancer.
\newblock {\em European Journal of Statistics}, 5:12--12.

\bibitem[Valdez and Pramanik, 2025b]{valdez2025exploring}
Valdez, I. and Pramanik, P. (2025b).
\newblock Exploring the interplay of adiposity, ethnicity, and hormone receptor
  profiles in breast cancer subtypes.
\newblock {\em arXiv preprint arXiv:2507.21348}.

\bibitem[Vikramdeo et~al., 2024a]{vikramdeo2024abstract}
Vikramdeo, K., Anand, S., Sudan, S., Pramanik, P., Singh, S., Godwin, A.,
  Singh, A., and Dasgupta, S. (2024a).
\newblock Abstract po3-16-05: Mitochondrial dna mutation detection in tumors
  and circulating extracellular vesicles of triple negative breast cancer
  patients for biomarker development.
\newblock {\em Cancer Research}, 84(9\_Supplement):PO3--16.

\bibitem[Vikramdeo et~al., 2024b]{vikramdeo2024mitochondrial}
Vikramdeo, K., Anand, S., Sudan, S., Pramanik, P., Singh, S., Godwin, A.,
  Singh, A., and Dasgupta, S. (2024b).
\newblock Mitochondrial dna mutation detection in tumors and circulating
  extracellular vesicles of triple negative breast cancer patients for
  biomarker development.
\newblock In {\em CANCER RESEARCH}, volume~84. AMER ASSOC CANCER RESEARCH 615
  CHESTNUT ST, 17TH FLOOR, PHILADELPHIA, PA~….

\bibitem[Vikramdeo et~al., 2023]{vikramdeo2023profiling}
Vikramdeo, K.~S., Anand, S., Sudan, S.~K., Pramanik, P., Singh, S., Godwin,
  A.~K., Singh, A.~P., and Dasgupta, S. (2023).
\newblock Profiling mitochondrial dna mutations in tumors and circulating
  extracellular vesicles of triple-negative breast cancer patients for
  potential biomarker development.
\newblock {\em FASEB BioAdvances}, 5(10):412.

\bibitem[Zang et~al., 2007]{ref11}
Zang, X., Thompson, R.~H., Al-Ahmadie, H.~A., Serio, A.~M., Reuter, V.~E., and
  Eastham, J.~A. (2007).
\newblock B7-h3 and b7x are highly expressed in human prostate cancer and
  associated with disease spread and poor outcome.
\newblock {\em Proceedings of the National Academy of Sciences},
  104:19458--19463.

\end{thebibliography}

\end{document}